# MeV Gamma-ray imaging spectroscopic observation for Galactic Centre and Cosmic Background MeV gammas by SMILE-2+ Balloon Experiment

Toru Tanimori  (on behalf of SMILE project )

Graduate School of Science, Kyoto University, Kyoto 606-8502, Japan

**Abstract**. Recently, there appears lots of papers on the possibility of light Dark Matter (DM) in MeV and sub-GeV scale.  Until now, only INTEGRAL and COMPTEL provided experimental data of 511keV of galactic center, and two spectra of Galactic Diffuse MeV gammas (GDMG)、and COMPTEL provided the Cosmic Background MeV gammas (CBMG) for wide sky for indirect detection of light DMs. However except 511keV, those spectra for diffuse gammas included large statistical and systematic errors in spite of 10 years observation, since both two instruments suffered from severe background radiation in space.  In 2018 April, we (SMILE-project in Comic-ray Group of Kyoto University) have observed MeV gamma rays for whole southern sky by Electron Tracking Compton Camera (ETCC) using JAXA balloon at Australia during one-day. (SMILE2+ Project)  By measuring all parameters of Compton scattering in every gamma, ETCC has achieved for the first time to obtain the complete direction of MeV gammas as same as optical telescopes, and also to distinguish signal gammas from huge background gammas in space clearly.  In this observation, ETCC with a large Field of View of 3sr observed MeV gammas from 3/5 of all sky including galactic centre, a half disk, crab, and most of CBMG By reconstructing the Compton process, we successfully obtained pure comic gammas by reducing background by more 2 orders, which is clearly certificated by the clear enhancement of detected gamma flux with ~30% during galactic center passing through the Field of View, which is consistent with the ratio of CBMG and GDMG. Now 511keV gammas GDMG are preliminarily detected with ~5 and >10σ respectively around Galactic Centre.  Also we have obtained near $10^5$ events of CBMG in with quite low background of only a few 10% in total CBMG events. Thus we obtained good data for both with high statistics and very low systematics even one day observation.

## 1. Introduction

Both the discovery of Dark Matter and the elucidation of nucleosynthesis are very important for comprehensive understanding Matter in Universe. For both, astronomical MeV γ observation is considered to be significant, which gives the direct observation of  nucleosynthesis fields such as supernova and material circulation, and also gammas from light DM annihilations. However, there have been only two uncertain data measured by COMPTEL [1] and INTEGRAL [2] due to unclear imaging method and very strong background which is mainly produced by the interaction of the instrument itself and cosmic rays. .In particular COMPTEL was a first Compton Camera used in Science [3]. So called "Compton Camera" (CC) reconstruct a Compton scattering process event by event and obtained the one of the two angles determining the direction of the incident gamma. since CC cannot measure the direction of the recoil electron in this process. Then CCs do not provide the direction of gamma completely as a point but a circle including its true direction as shown in Figures 1a and b Thus, CC cannot resolve the equation of the process completely, and it hardly reduce the backgrounds [4].

## 2. Principle and development of Electron Tracking Compton Camera (ETCC)

In order to overcome those problems, We (SMILE-project in Comic-ray Group of Kyoto University) have developed Electron Tracking Compton Camera (ETCC) consisting of a gas Time Projection Chamber (TPC) for tracking recoil electrons and GSO scintillator arrays (PSA) to explore MeV gamma astronomy since 2000 as shown in Figure.1a[5]. By measuring all tracks of Compton process generated in TPC, we completely reconstruct the direction of gammas as a point as shown in Figure 1b for all gammas scattering in TPC. We started the project of SMILE (Sub-MeV and MeV gamma ray Imaging Loaded-on balloon Experiment) for evaluating the ability of ETCC for exploring MeV gamma astronomy. In 2006, the first balloon borne experiment SMILE-I (SM1) with a small 10cm-cubic ETCC was done for the certification of its background rejection ability with observing a diffuse cosmic gamma-ray spectrum, where 420 cosmic gammas were successfully obtained from $\sim 10^5$ background by applying dE/dx cut for an electron track [5]. Then we developed a 30cm-cube ETCC as SMILE-2 (SM2) to detect gammas from celestial objects such as Crab [6], and attained the 100% detection of a fully contained event from 10% of S-I. Now two selections of dE/dx and fully contained track in TPC gives us a pure fully Compton events without losing Compton events[6] .A significant discovery is to define a well-defined Point Spread Function (PSF) in ETCC, which have make clear a reason of obscure imaging in conventional Compton Camera (CC)[4]. Such PSF dramatically reduce the background by 2-3 orders within PSF, which enabled us to estimate the sensitivity precisely similar to Optical or X-ray telescopes. When PSF is defined as half power radius, PSF is estimated from SPD (uncertainty of the direction of the recoil electron) and ARM (uncertainty of the direction of scattered gamma) in Figure.1a, and SPD of <10 degree with ARM of a few degree reaches the PSF of ~1 degree which gives an sensitivity of <1mCrab@$10^6$s with a few 100 cm$^2$ effective area [6]. Thus, ETCC is a unique MeV gamma telescope to do the proper imaging Spectroscopic observation which is essential method in all fields of astronomy for the first time in the world.

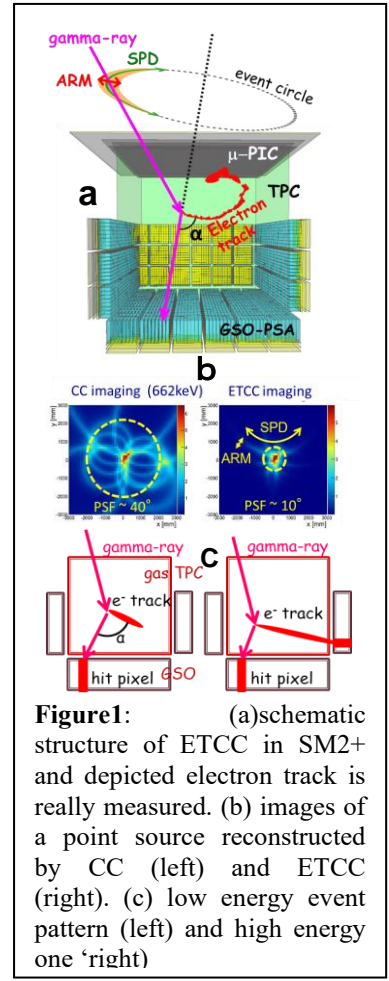

**Figure1**: (a)schematic structure of ETCC in SM2+ and depicted electron track is really measured. (b) images of a point source reconstructed by CC (left) and ETCC (right). (c) low energy event pattern (left) and high energy one 'right)

## 3. SMILE2+ balloon borne experiment

SMILE-2+ (SM2+) depicted in Figure.1a was improved from SM-2 for an enough sensitivity estimated based on this PSF for 511keV line gammas around Galactic center [7]. An effective area of SM2+ was increased 2 times at 511keV and >10 times above 1 MeV by pressuring Argon gas at 2atm and the use of a double thick GSO in the bottom Pixel Scintillation Arrays (PSA). Also since a fully contained electron in TPC restrict the energy range of gammas less 1MeV(low energy events), all PSAs were set in the TPC gas vessel to catch higher energy Compton recoil electrons penetrating from the gas TPC as shown in Figures 1a and 1c, which enables us to detect 1~7MeV gammas (high energy events).

We carried out one-day observation with a large Field of View (>3sr) for

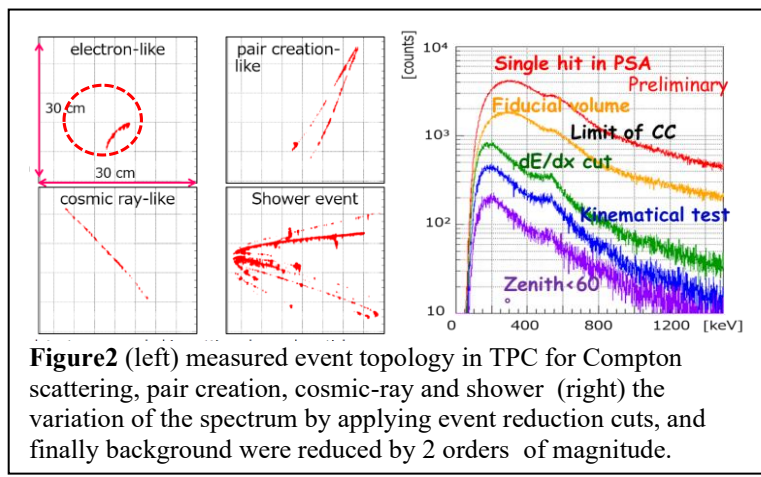

**Figure2** (left) measured event topology in TPC for Compton scattering, pair creation, cosmic-ray and shower (right) the variation of the spectrum by applying event reduction cuts, and finally background were reduced by 2 orders of magnitude.

whole southern sky by the JAXA balloon launching at Alice Springs of Australia in 2018 April 7~8. The balloon was successfully floated at the altitude of 37~39km during 30hours, in which we observed the southern sky including Galactic Center a half of Galactic disk, Crab, Cen-A and Sun as planned. Although analysis is under way, we successfully obtained pure comic gammas of $10^5$ events by reducing background by more 2 orders, as shown in Figure.2 right. Figure 2 left indicates the ability of the clear event topology measured by gas TPC for the identification of Compton Scattering, and Figure.2 right shows the variation of the spectra of events passing through the noise reduction cuts. The amount of final events is consistent with that estimated from SM1 results, and the dependence of the number of gammas on the altitude matches with the previous data well. Those strongly supports the signal to background ratio is larger than 1, which is also clearly certificated by the obvious enhancements of detected gamma flux with ~30% during Galactic Center passing through the Field of View (FOV), for both low and high energy events as shown in Figure.1c. This ratio is consistent with that previously reported data, and as a result, the final data

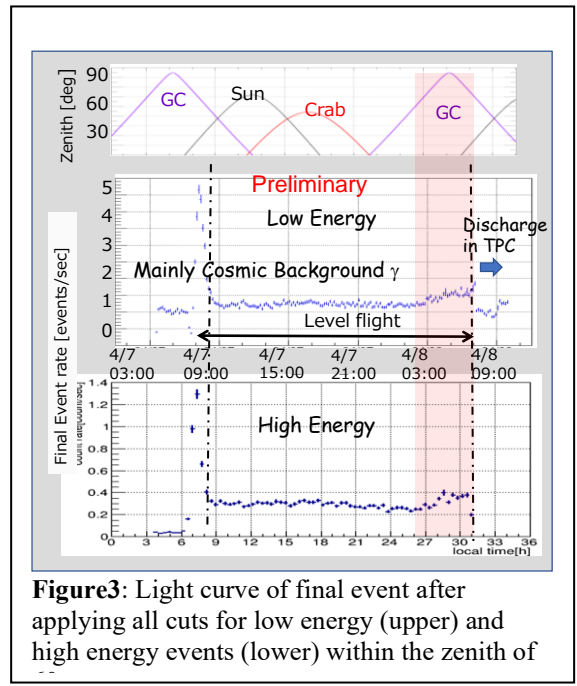

**Figure3**: Light curve of final event after applying all cuts for low energy (upper) and high energy events (lower) within the zenith of contains only a few 10% of background in the final event data. Thus, we have achieved the noise-free observation of MeV gamma-rays for the first time in the world. Until now, all observations of MEV gammas done before SM2+ includes about 2 order higher background than the signal.

To obtain the gamma-ray flux from GC, we set the ON circular region with the radius of 30° from GC since the enhancement in the light curve started about 2hours before GC reached the zenith. Off region was set the same radius including no Milky Way as shown in Figure.3. At present, we used only low energy events to estimate the intensity.. Figure 4 shows the law-count spectra of 3 ON regions (0.1~2MeV), OFF region and the residuals which responds to the spectrum of gamma from the ON region. An intense flux is observed from GC (ON-C), and weak one is from the ridge part (ON-A). No residual is observed from the outer region of Milky Way (ON-B). In this spectrum of GC 511keV line is clearly observed with 5σ only for 2hours observation. GDMG is detected more 10σ. These spectra include the effects of only the detection efficiency of the ETCC, but the effects of the most systematics error such as the absorption of the balloon payload and leakage from cosmic-rays were cancelled automatically by subtracting the spectrum of the OFF region. Then we corrected the detection efficiency by simulation study to obtain the actual intensity. Figure 5 is a preliminary result (0.2~1MeV) of the intensity from the 30 degree radial region around GC with the fluxes reported by

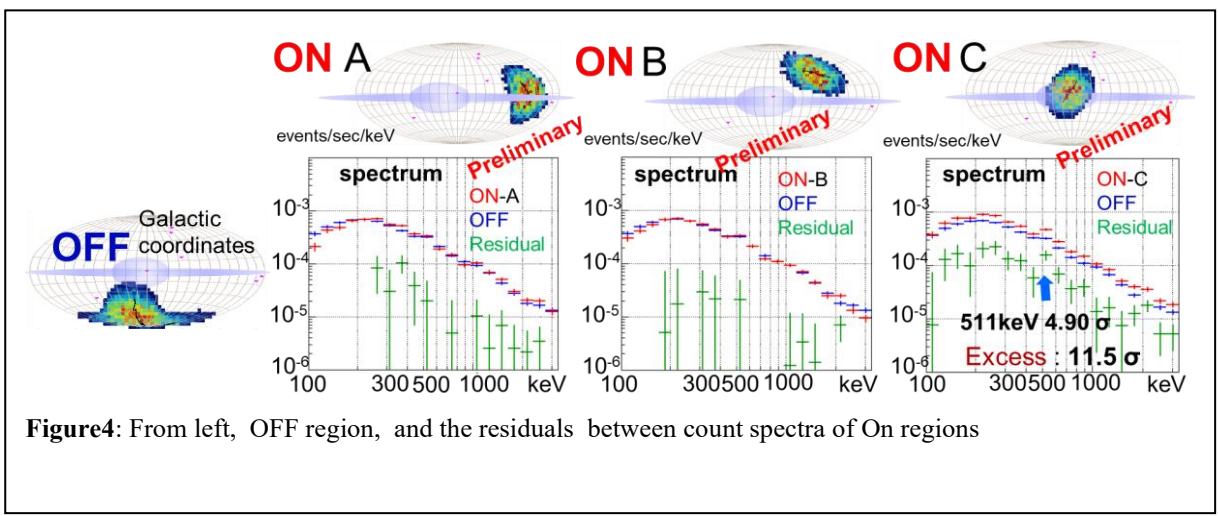

**Figure4**: From left, OFF region, and the residuals between count spectra of On regions

INTEGRAL (E<1MeV) [2] and COMPTEL (1<E<30MeV).[1]

## 4. Results and Conclusion

Our result looks consistent with that of INTEGRAL, and indicates near one order higher flux than theoretical estimation [8]. In sub-MeV and MeV regions, astronomical source of gammas is considered to be inverse Compton scattering by multi MeV and GeV cosmic electrons, and theoretical uncertainty for this calculation is considered to be relatively smaller than the estimation of GeV gamma [9]. Therefore, such a non-negligible discrepancy between the observation and theoretical estimation will be very interesting. As discussed in GeV region for Fermi data, DM will be one of interesting candidates. Now we are obtaining the higher intensity of 1~7MeV using higher energy events, and the intensity map. In particular, the intensity map will be a good indicator to search the origin of MeV gammas as shown in Figure. 6.

Thus, we have achieved the proper imaging spectroscopic observation of MeV gammas as same as that in X-ray and GeV gamma telescopes, and this success is certainly a keystone of the beginning of a new era for MeV gamma-ray astronomy.

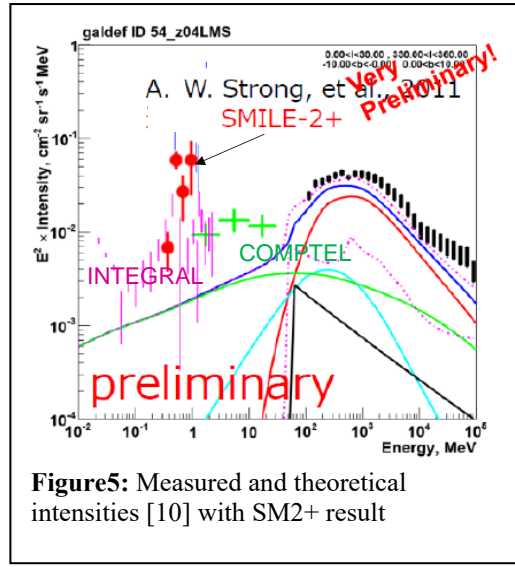

**Figure5:** Measured and theoretical intensities [10] with SM2+ result

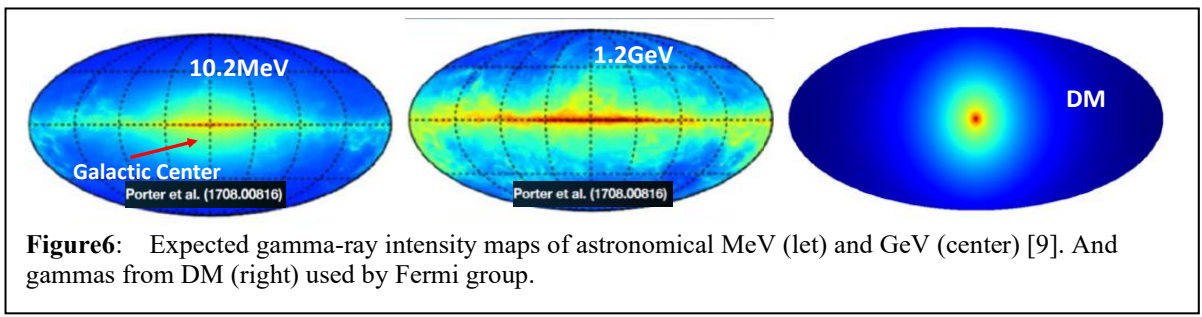

**Figure6**: Expected gamma-ray intensity maps of astronomical MeV (let) and GeV (center) [9]. And gammas from DM (right) used by Fermi group.